\documentclass[trackchanges]{aastex7} 

\usepackage{caption}
\usepackage{subcaption}
\usepackage{booktabs} 
\usepackage{footmisc}
\usepackage{tablefootnote}

\newcommand{\gray}{$\gamma$-ray}
\newcommand{\tgam}{$t_{\gamma}^{*}$}

\makeatletter
\let\old@foot@note@mark\footnotemark
\let\old@foot@note@text\footnotetext
\makeatother

\begin{document}

\title{An Optically Motivated Gamma-ray Study of Fermi-LAT Novae}
\shorttitle{Gamma-ray novae}

\author[orcid=0000-0002-0472-9195,sname='O. K. Henry']{Owen K. Henry}
\affiliation{The Graduate Center, Physics Program, City University of New York, 365 Fifth Ave., New York, NY 10016, USA}
\affiliation{Department of Astrophysics, American Museum of Natural History, Central Park West at 79th Street, New York, NY 10024, USA}
\email[show]{ohenry@amnh.org}  

\author[orcid=0000-0001-9139-0945, sname='Paglione']{Timothy A. D. Paglione}
\affiliation{Department of Earth \& Physical Sciences, York
College, City University of New York, 94-20 Guy R. Brewer Blvd., Jamaica, NY 11451, USA}
\affiliation{Department of Astrophysics, American Museum of Natural History, Central Park West at 79th Street, New York, NY 10024, USA}
\email{}
\author[orcid=0009-0008-4599-2935, sname='Zurek']{David Zurek}
\affiliation{Department of Astrophysics, American Museum of Natural History, Central Park West at 79th Street, New York, NY 10024, USA}
\email{}

\author[orcid=0000-0003-3835-8115, sname='Tan']{Joshua Tan}
\affiliation{Department of Natural Sciences, LaGuardia Community College, City University of New York, 31-10 Thomson Ave, Long Island City, NY 11101}
\affiliation{Department of Astrophysics, American Museum of Natural History, Central Park West at 79th Street, New York, NY 10024, USA}
\email{}

\begin{abstract}
High-energy (GeV) \gray \ emission from nova eruptions was a surprise detection by the Fermi-LAT in 2010. Since then, it has been suggested that the \gray s from these systems are generated in collisionless non-relativistic shocks.  In this theory, the observed correlation between optical and \gray \ nova lightcurves naturally arises if a portion of the optical luminosity is reprocessed shock power. In this work, we investigate this scenario by analyzing Fermi-LAT-detected novae in varied time bins and then correlating the bin size that maximizes the nova's significance (\tgam) with the nova's optical eruption data. Furthermore, we investigate whether there is a common optical decay across the sources' measured \tgam\ values in the population. We find that across our population, a nova's $t_3$ (the time it takes the nova V band brightness to decay 3 magnitudes) appears to be the favored analysis bin that optimizes the Fermi-LAT detection significance, although there is significant spread. Additionally, we report V679 Car, a source previously noted as a marginally detected \gray\ nova, as a $>5 \sigma$ detection in this work. We perform cross-correlation analysis of \gray\ and optical lightcurves.

\facilities{Fermi-LAT}

\software{Fermitools \citep{2019ascl.soft05011F}, 
          Fermipy \citep{2017ICRC...35..824W},
          scikit-learn \citep{scikit-learn},
          Astropy \citep{2013A&A...558A..33A,2018AJ....156..123A,2022ApJ...935..167A},
          numpy \citep{harris2020array},
          pandas \citep{reback2020pandas}
          }
\end{abstract}

\section{Introduction} 

Novae are celestial events that humans have observed for centuries. Since the first observations of novae, we have learned a great deal about these luminous transients; however, much remains unknown. Today, novae are classified into many different varieties, the two primary types being classical and embedded novae. Both classical and embedded novae originate from explosions on the surfaces of white dwarfs (WDs) in cataclysmic variable (CV) systems, but classical novae erupt from a CV with a WD and a main-sequence (MS) companion, while embedded systems have a post-MS companion and a relatively massive WD with a long-period orbit of $\sim 100$ days \citep{1995cvs..book.....W, 1997A&A...322..807D}. Symbiotic systems are a subset of embedded novae with slowly evolving eruptions \citep{1983ApJ...273..280K}. In a classical nova, the eruption is powered by a thermonuclear runaway (TNR) that occurs when enough hydrogen-rich material is accreted onto the WD from its MS companion \citep{1972ApJ...176..169S, starrfield_thermonuclear_2016}. A hydrogen shell forms around the WD, and eventually a runaway thermonuclear reaction begins, causing the system to erupt.

The Large Area Telescope (LAT) aboard the Fermi Gamma-ray Space Telescope discovered that both classes of nova were a surprisingly strong source of GeV \gray\ emission \citep{2010Sci...329..817A, 2013arXiv1308.6281H, 2019PhT....72k..38M}. In 2014, \citet{2014Sci...345..554A} (hereafter, the LAT collaboration) declared novae a new class of GeV
sources. TeV emission was theoretically predicted to originate from the nova
V407 Cyg \citep{2012PhRvD..86f3011S}, and was later detected from
the RS Oph system \citep{2022NatAs...6..689A, 2022Sci...376...77H, 2025A&A...695A.152A}. RS Oph remains the only nova from which TeV emission has been detected. Since 2014, studies have proliferated trying to understand the observed nova emission at high energies \citep{2019PhT....72k..38M, 2015MNRAS.450.2739M, franckowiak_search_2018}. 

Among the models developed to explain the GeV \gray s from novae, a leading theory suggests that the TNR powers an initial phase of  WD envelope expansion. The envelope is then ejected either by the TNR power or by the CV binary's orbital energy. If the envelope is ejected by the binary's orbital energy, this ejection is expected to take the form of a toroidal outflow confined to the WD's equatorial plane. 
This slow outflow is followed by a second ejection (or fast wind) from the WD with higher velocity and a more spherical geometry. The competing ejecta produce a bipolar outflow that lead to strong shocks where particles are accelerated by the Fermi process \citep{2013A&A...551A..37M, 2018ApJ...852...62V, chomiuk_new_2021, li_nova_2017, 2026arXiv260315480M, 2026ApJ..1004..109D}. Although it is possible that this multi-phase ejecta scenario is not relevant to symbiotic systems, since the ejecta would more likely interact with the dense wind of a massive companion \citep{2010Sci...329..817A, 2013A&A...551A..37M, 2015MNRAS.450.2739M}, in a study of RS Oph, \citet{2023ApJ...947...70D} suggested that the production of high-energy photons in symbiotic and classical systems might be more similar than suggested in the literature. They conclude this from spectral measurements of multiple velocity components from RS Oph, analogous to measurements of classical systems \citep{2020NatAs...4..776A}. Recently, \citet{2025arXiv251205220A} resolved multiple outflows in the V1674 Her and V1405 Cas systems from near-infrared CHARA\footnote{\url{https://www.mtwilson.edu/chara/}} images.

Some have argued that the optical luminosity of a nova event is powered by the same non-relativistic shocks that give rise to \gray s through absorption and re-emission of thermal X-rays \citep{2015MNRAS.450.2739M, 2015arXiv150606031M, 2017NatAs...1..697L,  2020NatAs...4..776A}. This scenario explains the lack of X-ray and radio emission early in the eruption due to the large bound-free and free-free optical depths \citep{2014MNRAS.437.1962H, 2014MNRAS.442..713M, 2025ApJ...988..211M}. Given the apparent close relationship between increases in optical and \gray\ flux, studies have attempted to investigate known optical tracers of shocks and properties of \gray\ emission \citep{franckowiak_search_2018, 2026MNRAS.546f2270C}. To explore how a nova's optical lightcurve can inform its \gray\ analysis, we correlate the optical and \gray\ lightcurves. Correlations of this kind have been previously reported \citep{2020NatAs...4..776A, 2026MNRAS.546f2270C}, and theoretically motivated (see, e.g., \citet{2015MNRAS.450.2739M, 2018ApJ...852...62V, 2017NatAs...1..697L}). As described above, X-rays get absorbed and re-radiated as optical photons, whereas the \gray s are promptly transmitted. The ``breakout" of a late X-ray signal from eruption events is a well-established phenomenon \citep{chomiuk_new_2021, 2025arXiv251220175H}. It is possible to fit for the X-ray absorbing column and determine the X-ray output before absorption. Studies have found that there typically is insufficient X-ray flux to account for \gray\ or optical emission, although solutions to this problem exist \citep[][and references therein]{2025ApJ...988..211M}.

In a recent study, \citet{2025arXiv251220175H} model the lightcurve of V1674 Her to investigate the late X-ray breakout observed in novae by closely inspecting its multi-wavelength emission components. In classical systems, where the shock is internal to multiple ejecta phases close to the WD surface, one might expect the \gray \ and optical lightcurves to be correlated with little lag. However, this expectation might not apply to all types of novae \citep{chomiuk_new_2021, 2015MNRAS.450.2739M, 2017NatAs...1..697L, 2020NatAs...4..776A}. For instance, in an embedded system, the shock might interact with a wind from a massive companion or a dense circumstellar environment, yielding \gray \ and optical lightcurves that would be correlated with a non-zero day lag given the more extended region of downstream material. While nova lightcurves are sufficiently similar to establish some families for convenience, the decay times across the population of observed novae span orders of magnitude in time \citep{2025ApJ...981..198C}. Investigating the correspondence between the \gray \ and optical lightcurves is essential for understanding the shock scenario of particle acceleration in nova environments as their primary \gray\ emission mechanism \citep{2020NatAs...4..776A}.  

We aim to test the correspondence between GeV and optical emission from novae.  
In the literature, the time it takes for a nova's optical light curve to decline by 2 magnitudes from maximum is called $t_2$, initially defined to parameterize a nova's ``speed class" \citep{1957gano.book.....G}. Various effects have been argued to influence these decay times, including distance, ejecta velocities, and maximum absolute magnitudes \citep{1995ApJ...452..704D, 2025ApJ...981..198C, 2018MNRAS.476.4162O}. Following this convention, we generalize the measurement to $t_N$ for $N$ magnitudes of decline. We developed a machine learning-based software tool in collaboration with the Legacy Survey of Space and Time Discovery Alliance (LSST-DA) Project Dovetail\footnote{\label{lsstda}\url{https://lsstdiscoveryalliance.org/introducing-project-dovetail/}} to allow for the $t_N$ decay times to be extracted directly from optical lightcurves (\S \ref{sec:tool}). Our hypothesis is that the optimal \gray\ time bin should correspond to a particular $t_N$ for novae,  given the physics of the eruption and emission mechanisms described above.

\section{Methods}\label{sec:methods}
\subsection{The Sample}\label{sec:sample}
We examine all 26 novae detected by Fermi-LAT from August 2008 through June 2024 on Koji Mukai's Fermi-LAT-detected novae list\footnote{\url{https://asd.gsfc.nasa.gov/Koji.Mukai/novae/latnovae.html} \label{list}}(hereafter, K. Mukai's list) with a typical \gray\ maximum likelihood analysis described in \S \ref{sec:like}. 
Of these sources, 11 are within $0.5^\circ$ of an unidentified 4FGL point source, and we investigate in this work whether any of these unidentified sources may actually be associated with the nova. With the large field of view of the LAT, the background must be properly modeled in order to ensure photons only associated with the target source are counted. The sources from the 4FGL catalog along with the diffuse Galactic and isotropic emission models were used as the initial background, which we refine by testing the \gray\ region of interest (ROI) with procedures described in \S \ref{sec:offtime} and \S \ref{sec:ontime}. If an unidentified Fermi-LAT source is in fact due to the nova of interest, then that source should be removed from the model background in subsequent analysis of the nova eruption so the recovered target flux is appropriate. 

All of the novae in our sample are presented in Table \ref{tab:fitin}. Eight nova eruptions that are cataloged in the 4FGL with the \texttt{NOV} source-class flag are in the first section of Table \ref{tab:fitin}.
The 11 novae with coincident unidentified 4FGL sources are listed in the second section of Table \ref{tab:fitin}. The third section of Table \ref{tab:fitin} has the \gray\ detected novae with no spatially coincident 4FGL counterpart.  
The optical data for this study are sourced from the American Association for Variable Star Observers (AAVSO)\footnote{\url{https://www.aavso.org}} and optical coordinates sourced from Bill Gray's \texttt{galnovae} list\footnote{\url{https://github.com/Bill-Gray/galnovae/blob/master/galnovae.txt} \label{gray}}. 

\begin{table}
\caption{Novae and spatially coincident 4FGL sources.}
\centering
    \begin{tabular}{@{} *4l @{}}    \toprule
\textbf{Nova / 4FGL}   & \textbf{Positional}  & \textbf{Optimal}    & \textbf{$t_{\gamma}^{*}$}\\
{\bf Counterpart}   & {\bf Offset ($^\circ$)} & {\bf ROI TS}\textsuperscript{a} &  {\bf (days)} \\
 \midrule
 V906 Car / 4FGL J1036.2-5936     & 0.35 & 3381.6 & 48.5 \\
 RS Oph / 4FGL J1750.3-0644       & 0.66 & 1927.0 & 27.4 \\
 V339 Del / 4FGL J2023.5+2046     & 0.54 & 394.6  & 15.5 \\ 
 V407 Cyg / 4FGL J2102.1+4546     & 0.36 & 513.1  & 15.5 \\
 V5856 Sgr / 4FGL J1820.8-2822    & 0.69 & 312.1  & 27.4 \\
 V1369 Cen / 4FGL J1353.3-5910    & 0.50 & 142.0  & 48.5 \\ 
 YZ Ret / 4FGL J0358.4-5446       & 0.22 & 88.2   & 8.7 \\
 V5668 Sgr / 4FGL J1837.6-2904    & 0.69 & 82.8   & 86.0 \\ 
 \midrule 
 FM Cir / 4FGL J1356.0-6747        & 0.46 & 307.5 & 846.7 \\
 V1723 Sco / 4FGL J1727.0-3835     & 0.44 & 191.0 & 15.5 \\
 \bf V1324 Sco / 4FGL J1750.9-3301 & \bf 0.40 & \bf 166.3  & \bf 48.5 \\ 
 \bf V5855 Sgr / 4FGL J1811.0-2725 & \bf 0.15 & \bf 99.5   & \bf 27.3 \\ 
 V1716 Sco / 4FGL J1723.6-4126     & 0.23 & 56.5   & 15.5 \\
 V679 Car / 4FGL J1115.1-6118      & 0.16 & 61.97  & 269.8 \\ 
 V1707 Sco / 4FGL J1736.9-3525     & 0.26 & 26.3   & 4.9 \\
 V3890 Sgr / 4FGL J1830.7-2414     & 0.23 & 54.4   & 152.3 \\
 \bf V549 Vel /  4FGL J0851.2-4737 & \bf 0.18 & \bf 38.4 & \bf 86.0 \\
 V1405 Cas / 4FGL J2326.5+6122     & 0.28 & 13.6   & 86.0 \\
 V745 Sco / 4FGL J1755.7-3259      & 0.27 & 11.2   & 4.9 \\ 
 \midrule  
 V959 Mon  / NA                   & -- & 182.8  &  4.9  \\
 V392 Per  / NA                   & -- & 124.0  &  8.7 \\
 V357 Mus  / NA                   & -- & 75.2   & 27.4 \\
 V1674 Her / NA                   & -- & 30.6   &  1.6  \\
 V6598 Sgr / NA                   & -- & 18.2   &  15.5 \\
 V1535 Sco / NA                   & -- & 16.9   &  8.7  \\
 V407 Lup  / NA                   & -- & 14.4   &  4.9  \\
 \bottomrule
 \hline
 \multicolumn{4}{@{}l}{a) ROI model determined with the tests in \S \ref{sec:offtime}, \S \ref{sec:ontime}.}
  \\
\end{tabular}
\tablenotetext{}{For novae with a 4FGL counterpart, positional offsets are measured between the 4FGL source and the optical position of the 
nova from \footnotemark[\getrefnumber{gray}]. 
The time bin that maximizes TS in the ROI (\tgam) is shown in the last column. The table is separated into three samples: the top are the novae in the 4FGL catalog, the middle are the novae with a 4FGL source within $0.5^{\circ}$ of their optical position, the bottom are novae with no nearby 4FGL point source. The sources in bold are potential associations with the 4FGL catalog and we recommend that they be removed from the ROI model during the eruption window.} 
\label{tab:fitin}
\end{table}

\subsection{Binned Likelihood Analysis}\label{sec:like}
We adopt the standard maximum likelihood analysis to search for \gray\ emission from our target novae \citep{1996ApJ...461..396M}. The aim of this analysis is to discern whether or not there is a \gray\ emitting source at each target nova's coordinate through a likelihood test.
We filter the data using a zenith-angle cut of $90^{\circ}$ to avoid terrestrial contamination and choose a photon energy range of 50 MeV to 300 GeV, 
split into 30 logarithmically spaced bins\footnote{\url{https: //fermi.gsfc.nasa.gov/ssc/data/analysis/scitools/} and \url{http://fermipy.readthedocs.io/en/latest/quickstart.html}\label{fssc}}. This energy range has been shown to maximize the sensitivity of novae analysis with Fermi-LAT \citep{collaboration_fermi_2014}. The ROIs are $21^{\circ}\times 21^{\circ}$ on-sky analysis zones, centered around each target coordinate. The photon event data are collected, binned, and fit by the maximum likelihood analysis within the \gray\ ROI. Its size accounts for the large field of view and PSF of the Fermi-LAT. Catalog sources an additional 10\degr\ outside the ROI are added to the analysis to account for photons from sources (especially bright and variable sources) farther from the ROI center but which can still contribute photons to the target location \footnotemark[\getrefnumber{fssc}]. We use the third revision of the Pass 8 (P8R3) instrument response function (P8R3\_SOURCE\_V3), the most recent Galactic emission model (\texttt{gll\_iem\_v07.fits}), and isotropic background emission model (\texttt{iso\_P8R3\_SOURCE\_V3\_v1.txt}) \citep{2009ApJ...703.1249A} with the default event class and type (evclass = 128, evtype = 3). 
We separate the data into four point spread function (PSF) components (PSF0, PSF1, PSF2, PSF3) to match the response more accurately by PSF type.

We perform this analysis using \texttt{fermipy}, a Python package that facilitates analysis of LAT data with the Fermi Science Tools within the open-source Python distribution, Mamba\footnote{\url{https://github.com/mamba-org/mamba}} \citep{2019ascl.soft05011F, 2017ICRC...35..824W}. We perform the maximum likelihood test for the presence of a \gray\ point source at each target's location on the sky.  The result of the likelihood analysis is the Test Statistic (TS), defined as TS $ = 2\text{ln}(L/L_0)$, where $L$ is the likelihood of a point source being present at the center of the ROI, and $L_0$ is the null hypothesis that there is no central source \citep{1996ApJ...461..396M}. The detection significance can be estimated from $\sqrt{\textrm{TS}}$, and we adopt the usual detection threshold of TS $>25$ \citep{2022yCat.9067....0A}. 

\subsection{Off-peak Analysis}\label{sec:offtime}
To confirm the best ROI model, especially given the 11 coincident unidentified catalog objects, we determine the significance of the coincident 4FGL objects when the nova is not erupting. If the 4FGL source remains significant outside the nova eruption window (TS $> 25$), we retain it in the background model with the nova in the ROI. If the 4FGL source TS value diminishes outside the eruption window, we remove it from the model when fitting the nova.

We use the 16-year data set retrieved from the LAT data server.\footnote{\url{https://fermi.gsfc.nasa.gov/cgi-bin/ssc/LAT/LATDataQuery.cgi}} We then remove the photon files corresponding to the Fermi mission week in which the eruption was discovered, plus an additional 26 weeks afterward, and  
a buffer week before the eruption event. The long exclusion period ensures we minimize the potential contribution of the nova event to the binned likelihood analysis. 
We perform the analysis at the cataloged position of the 4FGL source and compare the resultant TS value to that measured over the entire Fermi mission. 

\subsection{On-peak Time Analyses}\label{sec:ontime}
We further evaluate the ROI models for each nova, particularly for any source that could be associated with an unidentified 4FGL object. For one, we center the ROI at the 4FGL source position and fit it using the standard analysis tools. For the second,  we fit two sources in the ROI: at the 4FGL catalog source position and at the nova's optical position. Finally, we fit a single source at the nova's optical position. 
For all analyses, spectral parameters of 4FGL sources within $5^{\circ}$ of the center of the ROI are free to be fit, and those outside remain fixed. For the central sources, we find generally that the power law with exponential cutoff (PLEC) 
spectral model yields higher or equivalent detection significance: 
\begin{equation}
    \frac{dN}{dE} = N_0(\frac{E}{E_0})^{\Gamma}e^{-({E}/{E_c})^b},
\end{equation}

where $N_0$ is the normalization, $\Gamma$ is the power law index, $E_0$ is an energy scale factor, $E_c$ is the cutoff energy, and $b$ is a second power law index that determines the curvature at the cutoff. The PLEC spectral model is typically favored for novae \citep{2014Sci...345..554A, collaboration_fermi_2014}. For all spectral fits, the cutoff energy, $E_c$, was fixed at 1 GeV, and the second power law index, $b$, was fixed at 1. We proceed with the PLEC model with the power-law index and normalization free to be fit for all sources in this work for consistency. 

We show the maximum TS for the most significant ROI model in (Table~\ref{tab:fitin}). We confirm any claim from \S \ref{sec:offtime} if the ROI model with the nova at its optical position maximizes TS.

For all 26 novae on K. Mukai's list, we explored a range of time bin sizes ($t_{\gamma}$) for analysis. 
The array of bin sizes starts at a minimum of $0.5$ days to ensure sufficient photon counts for the analysis, and ranges up to 1500 days, sampled logarithmically. 
For each source, we conduct a binned likelihood analysis for each $t_{\gamma}$. Each analysis is altered by modifying the \texttt{tmin} and \texttt{tmax} keywords in the \texttt{fermipy} configuration file, where \texttt{tmin} is the \gray\ discovery date from K. Mukai's list. If a time range is provided for the discovery, the earliest time is used as the eruption start time. We corroborate these times with Koji Mukai's List of Recent Galactic Novae\footnote{\url{https://asd.gsfc.nasa.gov/Koji.Mukai/novae/novae.html}}. However, there are exceptions. 
For most sources, the \gray\ outburst discovery date coincides reasonably with the start of the AAVSO data, but in two cases (V906 Car and V959 Mon) the times differ. 
We take the final time as $\texttt{tmax} = \texttt{tmin} + t_{\gamma}$. We define \tgam \ as the time bin that maximizes the TS. 
%
%
V906 Car initially erupted during the Fermi downtime in March-April 2018, which was due to a malfunctioning solar panel aboard the spacecraft, as noted by \citet{2020NatAs...4..776A}. In this case, the AAVSO data predate the \gray\ signal by a month. We take \texttt{tmin} to be the optical eruption start date, 2018-03-22, but can only conduct the test described in \S \ref{sec:ontime} for $t_\gamma \geq 27.39$ d. This source is not included in Fig. \ref{fig:violins}, nor is it included in the analysis described in \S \ref{sec:xcorr}. 
%
The \gray\ data precede the 
AAVSO data for V959 Mon by approximately 3 months because the nova was up during the day when the \gray\ eruption began. We take \texttt{tmin} for this source to be the earlier date on K. Mukai's list. This source is not included in Fig. \ref{fig:hist}, Fig. \ref{fig:violins}, or the analysis described in \S \ref{sec:xcorr}.

\begin{figure}
    \centering
    \includegraphics[width=0.8\linewidth]{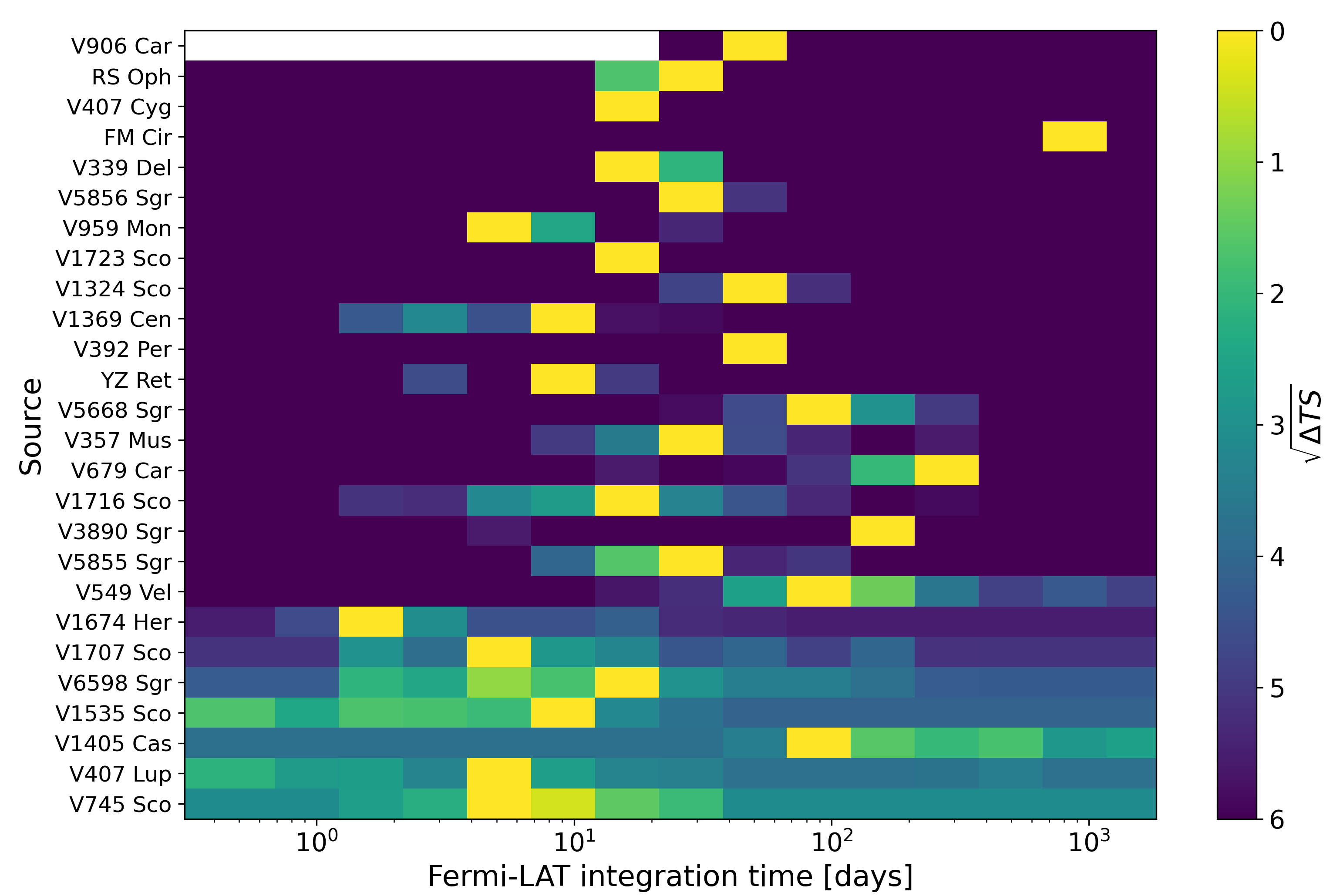}
    \caption{The results of the on-peak time analysis described in \S \ref{sec:ontime}. The color bar is mapped to the $\sqrt{\Delta TS} \sim \sigma$.  where $\Delta$TS is the difference between the maximum TS value in the optimal time bin (\tgam) and TS for the other integration times. The novae are in descending order according to their maximum TS, with the most significantly measured novae at the top. The first 7 analysis bins for V906 Car are white because the binned likelihood analysis could not be compiled during the Fermi downtime (see text
    ).} 
    \label{fig:cmap}
\end{figure}

\section{Results}\label{sec:results}

Measured \tgam \ and TS for each source are shown in Table \ref{tab:fitin}. 
We illustrate the results of this analysis in Fig. \ref{fig:cmap}. In this work, we recover their high \gray\ significance.       
The color scale is mapped to $\sqrt{\Delta {\rm TS}}$, where $\Delta$TS is the difference between the maximum TS value from the fit in the \tgam \ window, and the TS value for other $t_{\gamma}$ bin sizes. We interpret $\sqrt{\Delta {\rm TS}}$ as a $\sigma$ from this maximum \citep{wilks1938}.

The novae are shown in descending order according to their maximum TS. For the highest TS novae, the \tgam \ time bin yields a much more significant fit than the others, sharply peaking at \tgam. Conversely, less well-detected novae show broad time bin distributions supporting a range of comparably significant $t_{\gamma}$ values. The eight novae cataloged in the 4FGL appropriately occupy the top. Marginally detected novae towards the bottom third of the plot have broader TS distributions with few time bins exceeding $3 \sigma$ from their maxima. While our \tgam\ measurements largely agree with the analysis conducted in \citet{franckowiak_search_2018}, in several cases, our measured \tgam \ differ from their claim of a common 15-day analysis window across the entire population of novae. For V1723 Sco and V6598 Sgr, our \tgam\ values agree with those published recently in \citet{fauverge2025fermilatdetectionsnovaev1723}. Likewise, this work recovers the lag and decay measurements for V1674 Her and V1405 Cas, recently reported in \citet{2025arXiv251220175H, 2025arXiv251205220A}. Lastly, we recover the marginal significance of the sources described as ``Possible Detection'' in K. Mukai's list. Four of these marginally detected sources, V6598 Sgr, V407 Lup, V1535 Sco, and V745 Sco, 
have a \gray\ significance at the $2 \sigma$ level in this study. One source, V679 Car, with our windowing, is measured to have a $>5 \sigma$ \gray\ significance. We also note that V1324 Sco, V5855 Sgr, and V549 Vel have nearby 4FGL sources that are insignificant outside the eruption window. We consider these potential associations with the 4FGL catalog and recommend that they be removed from the ROI model during the eruption window. We highlight these sources in bold in Table \ref{tab:fitin}. 

\section{Discussion}\label{sec:disc}

\subsection{$t_N$ Measuring Tool}\label{sec:tool}
To probe for correspondence between \gray \ and optical nova lightcurves, we developed an open-source tool (hereafter, \texttt{nova-times}\footnote{\url{https://github.com/project-dovetail/nova-times}}) for measuring novae lightcurves from AAVSO in collaboration with the LSST-DA\footnotemark[\getrefnumber{lsstda}]. In the literature, various methods have been used to measure nova decay times, such as linear interpolations \citep{2026MNRAS.546f2270C} or simply by-eye \citep{1985ApJ...292...90C}, which can introduce errors on the order of days  \citep{1993A&A...277..103C}. Since these quantities are vital for nova speed classifications, measurements of average absolute magnitudes, and scaling relations \citep{1993A&A...277..103C, 1995ApJ...452..704D, 2025arXiv251220175H, 2018MNRAS.476.4162O}, we developed a versatile tool to compute them. 

The \texttt{nova-times} tool takes in lightcurve data and returns a $t_N$, where $N$ is the number of magnitudes the nova has decayed from maximum and $t_N$ is the time over which this decay takes place. The tool uses a gradient boosted machine (GBM), a machine learning method, to infer data from a dataset with gaps, such as irregularly sampled lightcurves. The model uses weak learners (e.g., a tree method) to fit our time-series data in sequential learning epochs. At each iteration, it computes the mean squared error (MSE) and adjusts the fit to minimize the MSE. This corresponds to moving along the gradient of the loss function (which can be the MSE or another loss), which gives the model its name \citep{2023mlpa.book.....A}. The tool also offers options for simpler fitting techniques, such as interpolation. 

The tool accepts lightcurve data in any format from the AAVSO data server\footnote{\url{https://www.aavso.org/data-download}}. 
Presently, we have three methods to measure lightcurves: a simple ``nearest point" finder, a univariate spline \citep{Di1975}, and a GBM. For the GBM, we tune a data richness threshold at which the GBM depth increases to prevent overfitting while preserving the method's robustness. Detailed documentation is available at its GitHub repository. We report $t_2$, $t_3$, and $t_4$ measured with the \texttt{nova-times} program using V band lightcurves for the sources investigated in this work in Table \ref{tab:t234}. Our measured optical decays are comparable to those in literature \citep{2026MNRAS.546f2270C}. It should be noted that AAVSO lightcurves may not be properly calibrated, and peak eruption may be missed. 
The \texttt{nova-times} tool does not predict the peak brightness, so we assume that the observations sample the peak brightness of the nova. 

When testing the tool, we find that for sparser data, as expected, the variance in $t_N$ measurements systematically increases. We encourage users of {\tt nova-times} to always inspect their lightcurve data and fitting results with the visualization package ({\tt viz}) included in {\tt nova-times} to ensure the accuracy of measured lightcurve decays. We use the GBM fit for all decay measurements in this work. 

\begin{figure}
    \centering
    \includegraphics[width=0.8\linewidth]{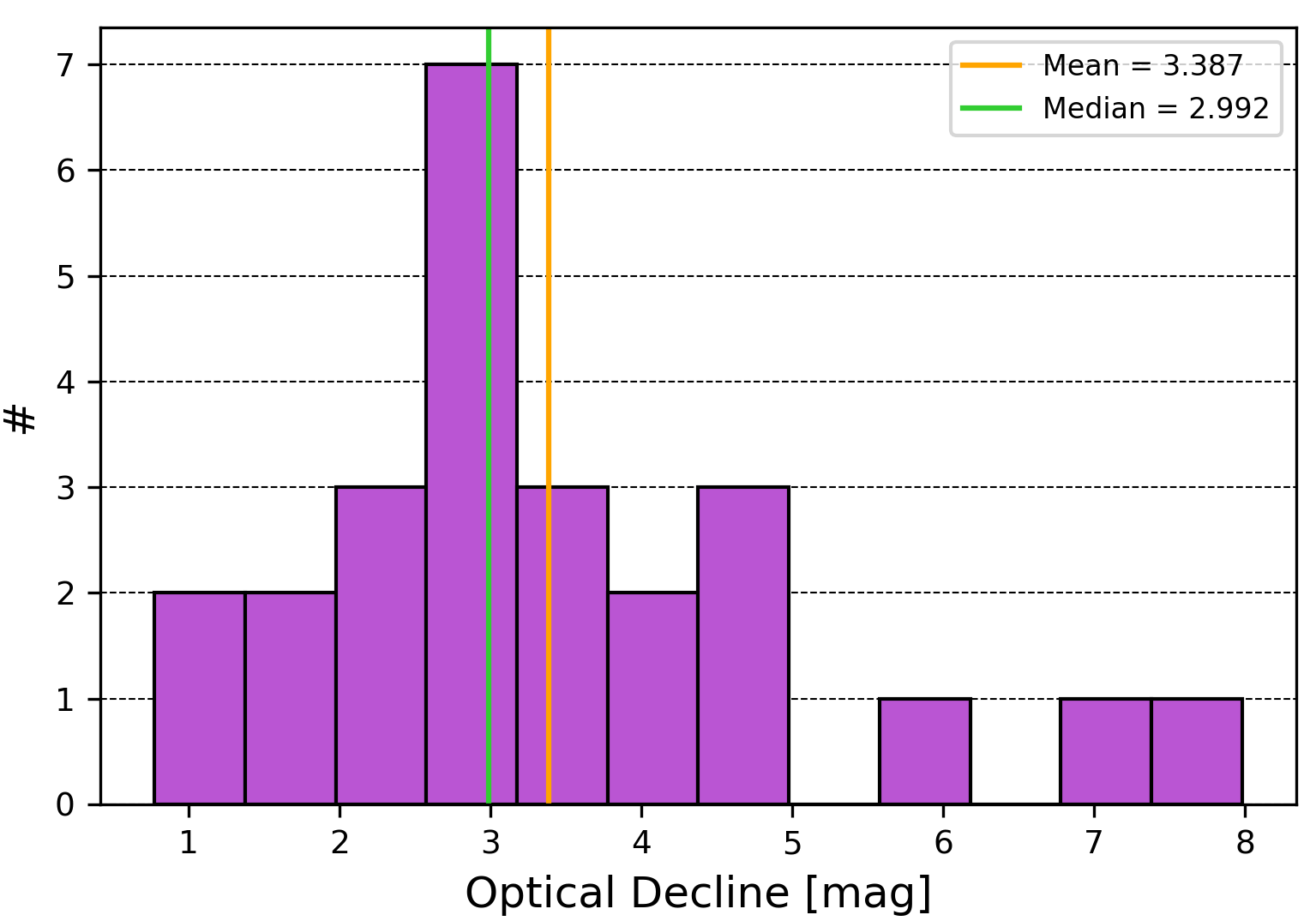}
    \caption{Histogram of the calculated optical magnitude drop corresponding to \tgam \ using the {\tt nova-times} tool. We can compare \tgam \ to the optical magnitude decrement in that same time from the AAVSO lightcurve. 
    The distribution peaks for a brightness drop of $\sim 3$ magnitudes, but the spread is substantial. We investigate this spread in \S \ref{sec:corr} and display the results in Fig. \ref{fig:violins}.}
    \label{fig:hist}
\end{figure}

\subsection{Optical Magnitude Decay Correspondence}\label{sec:corr}
\begin{figure}
    \centering
    \includegraphics[width=0.9\linewidth]{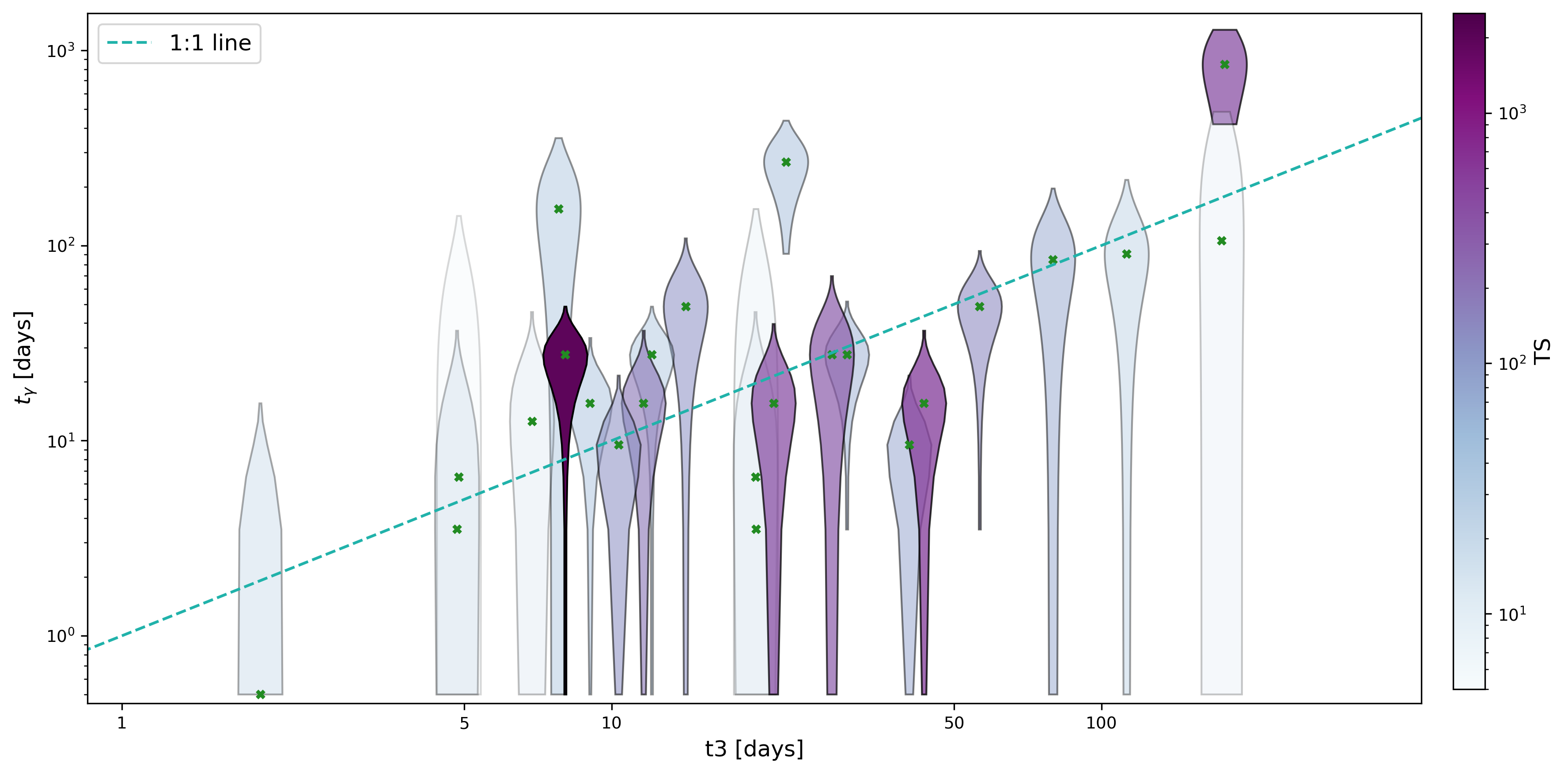}
    \caption{Plot of $t_\gamma$ from Fig. \ref{fig:cmap} against $t_3$, with the width of the regions determined from $\sqrt{\Delta{\rm TS}}$. The optimal time bin, \tgam, is indicated by the green cross. The regions are color-coded according to their maximum TS. The 1:1 line is included to guide the eye. 
    }
    \label{fig:violins}
\end{figure}

Here we expand on the result depicted in Fig. \ref{fig:cmap}. The test described in \S \ref{sec:ontime} yields a measure of TS for each $t_{\gamma}$. Plotting TS as a function of $t_{\gamma}$ results in a ``likelihood curve." The likelihood curves for each source are converted into a $\sqrt{\Delta \text{TS}}$ between the source's maximum TS value, TS(\tgam), and the rest of the nova's likelihood curve. 
We attempt to directly compare \tgam \ measured with the methods described in \S \ref{sec:ontime} with the nova's change in magnitude, measured from the magnitude at \texttt{tmin} to that at \texttt{tmax} according to the window, \tgam .  
Using this optimal integration time, we measure the $t_N$ with the tool described in \S \ref{sec:tool}. Fig. \ref{fig:hist} shows these results, which center on $N\approx 3$, indicating that $t_3$ is often the optimal integration window for the \gray\ data. The first and third quartiles of the $t_N$ distribution are $N \approx$ 2 and 4, respectively. We plot $t_{\gamma}$ against the measured decay time corresponding to a magnitude drop of $N=3$ in Fig. \ref{fig:violins}. Many sources lie on the 1:1 line in this figure, with some spread. As in Fig. \ref{fig:cmap}, the best-detected sources in this population (namely, the eight already cataloged in 4FGL-DR4) are more localized. The undetected sources have significance that is spread over more of the $t_\gamma$ domain. The analogous plots for $t_2$ and $t_4$ are shown in the appendix (Fig. \ref{fig:24violins}). We find that $t_3$ and $t_4$ have similar root mean square distances between their respective maxima and the 1:1 line, while the $t_2$ results are farther from the 1:1 line.

\subsection{Light curve cross-correlation}\label{sec:xcorr}

We also perform a cross-correlation study of the optical and \gray\ lightcurves to test whether the optical luminosity of novae is powered by the same shock environments that give rise to GeV \gray s. 
For all sources with TS$\: >30$ (18 novae) we create 25-bin \gray\ lightcurves within the sources' measured \tgam .  To more confidently report lags between the optical and \gray\ light curves, we cross-correlate the light curves, their first order finite differences with respect to time, and their second order finite differences with respect to time. We report any measurement that agrees across all three correlations and exceeds the time step ($\Delta t = t_{\gamma}^{*}/25$). 

We recover the anticipated lag of 0 days for most novae, albeit with considerable uncertainty. For RS~Oph, V1723~Sco, V357~Mus, and V5856~Sgr, we measure lags of $\sim 1$ day (all within each of their respective $\Delta t$ uncertainties). These sources might deviate from the zero-lag expectation due to fundamental differences in their system makeup. For example, RS Oph is a symbiotic nova with a red giant companion, which differs from more common classical novae. It would take longer for the ejecta to interact with this more extended circumstellar environment, where the nova ejecta might be shocking the diffuse wind from the binary companion instead of multiple phases of its own ejecta. \citet{cheung_fermi_2022} measure a comparable lag for RS Oph ($\sim 1.3$ days). 

Another interesting case is V1674 Her, which is designated as an intermediate polar, and is the fastest nova to date with a $t_2 \sim 0.9$ days \citep{2024BAAA...65...60L, 2025arXiv251220175H}. This work recovers the lag and decay results for V1674 Her (and V1405 Cas) recently reported in \citet{2025arXiv251220175H, 2025arXiv251205220A}. For V1674 Her, \citet{2025arXiv251220175H} suggests that the nova is the first to show a GeV \gray\ peak that substantially precedes the optical maximum. The first order finite difference cross-correlation test yields an optical lag of 4.56 hours, but the other two tests show no lag. Our optical decay measurements are consistent with the very short $t_2$ reported in their work (Table \ref{tab:t234}).

\section{Conclusions}\label{sec:conc}
We studied all 26 nova eruptions in K. Mukai's list of Fermi-LAT-detected novae from the Fermi Launch in 2008 until June 2024. We investigated 11 sources within a $0.5^{\circ}$ of an unassociated point source in the 4FGL-DR4. Based on the lack of significance outside the eruption window, three sources (V1324 Sco, V5855 Sgr, and V549 Vel) should be associated with a coincident 4FGL source. Therefore, these 4FGL sources (4FGL J1750.9-3301, 4FGL J1811.0-2725, 4FGL J0851.2-4737) should be removed from the model for the likelihood analysis.

We present an open-source code (\texttt{nova-times}) for measuring the decay times of optical transients, developed with LSST-DA. We use \texttt{nova-times} to impute data from sparse lightcurves to obtain more accurate measurements of lightcurve decay times, characterized by $ t_N$, which are typically used to classify novae and understand their physical properties.

The detection of GeV \gray s from nova systems by the Fermi-LAT suggests the presence of non-thermal particle acceleration in circumstellar shocks, analogous to cosmic-ray generation in supernova remnants. We investigate this physical scenario by identifying how the optical light curves of Fermi-LAT-detected novae can be used to optimize \gray\ detection significance. This is physically motivated by the theory of non-relativistic shocks internal to nova ejecta. In the literature, there is an apparent temporal coincidence between \gray\ and optical outbursts. The optical/\gray\ correspondence is explained by the optical photons being generated in the same shock events that accelerate the particles, which give rise to \gray s. In this work, we find that the optimized window to observe a Fermi-LAT nova eruption (\tgam) varies widely across the population. We relate the \tgam \ measured with the on-peak test with the optical magnitude decay for all \gray \ novae. Given the variety of optical nova lightcurves, we propose that the optimal analysis window can be based on the optical lightcurve, supported by tight correlations we measure between the optical and \gray\ lightcurves.

\begin{acknowledgments}
    The authors would like to thank LSST-DA for providing software engineering support through their Project Dovetail program. In particular, OKH would like to profusely express gratitude to Dr. Destry Saul of the LSST-DA Project Dovetail, who generously donated his time to spearhead the development of the \texttt{nova-times} tool. OKH would like to thank the AAVSO, particularly Dr. Bert Pablo and Sara Beck, for their prompt responses to queries regarding data availability. OKH thanks Dr. Koji Mukai for their meaningful exchanges and for diligently maintaining his Fermi-LAT-detected nova list. Similarly, OKH thanks Dr. Bill Gray for rapidly updating his list of Galactic Novae. This work was supported in part by the National Science Foundation under award \# AST-2219090, by the NASA New York Space Grant Consortium (awards \#80NSSC20M0096 and \#80NSSC25M7095), and by the Simons Foundation under award \# 00533845. This project made use of computational systems and network services at the American Museum of Natural History, supported by the National Science Foundation via Campus Cyberinfrastructure Grant Award \# 1827153 (CC* Networking Infrastructure: High-Performance Research Data Infrastructure at the American Museum of Natural History). The authors would also like to thank the referee of this publication for their thoughtful insight into this study.    
    
\end{acknowledgments}

\pagebreak 

\appendix

\section{Correlation between \tgam \ and optical decay time}

\begin{figure}[h]
    \centering
    \includegraphics[width=0.9\linewidth]{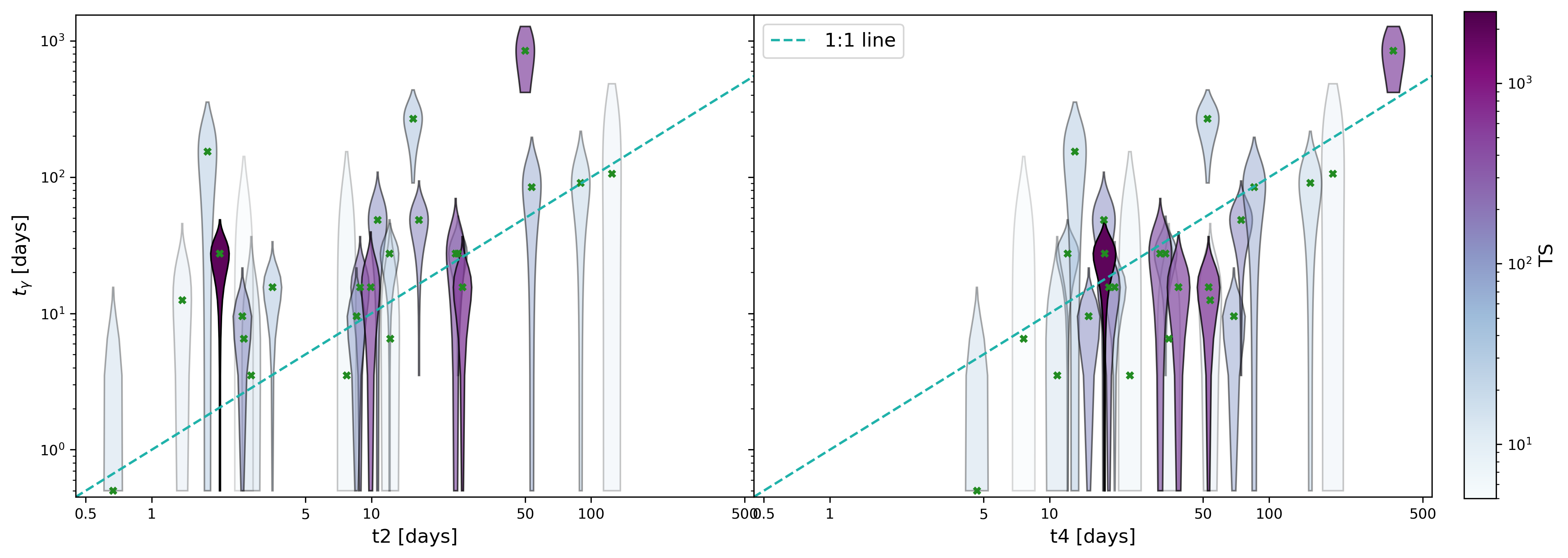}
    \caption{Distributions of $t_2$ and $t_4$ similar to the one for $t_3$ in Fig. \ref{fig:violins}.}
    \label{fig:24violins}
\end{figure}

In Fig. \ref{fig:24violins}, we present the distributions of $t_2$ and $t_4$ from the analysis we perform in \S \ref{sec:corr} for $t_3$. We present the $t_2, t_3, \: \text{and} \: t_4$ measurements from the \texttt{nova-times} tool (\S\ref{sec:tool}) in Table \ref{tab:t234}.

\begin{table}
\centering
\caption{Measured V-band decay times measured with the \texttt{nova-times} tool described in \S \ref{sec:tool}. All entries in this table are measured in days from the eruption maximum. It is presented in ascending order of $t_2$.}

\begin{tabular}{lccc}
\toprule
\textbf{Nova ID} & \textbf{$t_2$} & \textbf{$t_3$} & \textbf{$t_4$} \\
\midrule
V1674 Her & 0.7 & 1.9 & 4.7 \\
V6598 Sgr & 1.4 & 6.9 & 53.9 \\
V3890 Sgr & 1.8 & 7.8 & 13.0 \\
RS Oph & 2.0 & 8.0 & 17.8 \\
V392 Per & 2.6 & 10.3 & 15.1 \\
V745 Sco & 2.6 & 4.9 & 7.6 \\
V1707 Sco & 2.8 & 4.8 & 10.8 \\
V1716 Sco & 3.5 & 9.0 & 19.8 \\
V407 Lup & 7.7 & 19.7 & 23.2 \\
YZ Ret & 8.5 & 40.5 & 69.0 \\
V1723 Sco & 8.9 & 11.6 & 18.6 \\
V339 Del & 9.9 & 21.4 & 38.7 \\
V1324 Sco & 10.7 & 14.2 & 17.7 \\
V5855 Sgr & 12.1 & 12.1 & 12.1 \\
V1535 Sco & 12.2 & 19.7 & 34.9 \\
V679 Car & 15.5 & 22.7 & 52.5 \\
V1369 Cen & 16.5 & 56.5 & 74.5 \\
V5856 Sgr & 24.2 & 28.2 & 31.9 \\
V357 Mus & 24.7 & 30.2 & 33.7 \\
V407 Cyg & 26.0 & 43.5 & 53.0 \\
V906 Car & 41.2 & 59.4 & 100.2 \\
FM Cir & 50.1 & 178.6 & 367.6 \\
V5668 Sgr & 53.7 & 79.7 & 85.5 \\
V549 Vel & 89.6 & 112.7 & 153.9 \\
V1405 Cas & 124.3 & 176.1 & 195.1 \\
V959 Mon & 154.0 & 227.0 & 385.2 \\
\bottomrule
\end{tabular}
\label{tab:t234}
\end{table}

\pagebreak
\bibliography{sample7}{}
\bibliographystyle{aasjournalv7}

\end{document}